\documentclass[preprint,12pt]{elsarticle}
\usepackage{amssymb}
\usepackage{url}

\usepackage{amssymb}
\usepackage{afterpage}
\usepackage[section]{placeins}

\journal{Elsevier Science}

\begin{document}

\begin{frontmatter}

\title{Finite Lattice and Phenomenological Approximations for the Anomaly in the Density of a Water-like Lattice Gas Model}

\author{M.R. Thielo }
\ead{thielo@if.ufrgs.br}
%
\author{M.C.B. Barbosa}
\address{Instituto de F\'{\i}sica - Universidade Federal do Rio Grande do Sul, Caixa Postal 15051, \\ 91501-970 Porto Alegre, Rio Grande do Sul, Brazil}
\ead{barbosa@if.ufrgs.br}

\begin {abstract} We propose a model for a two dimensional, associative water-like lattice gas with one single variable representing both long and short-range interactions. The corresponding hamiltonian was solved exactly, by state enumeration in a finite lattice, so to obtain an analytic expression for the partition function. The lattice dimensions were chosen based on geometric characteristics of the stable phases found in previous works using Monte Carlo simulations. An expression for the particle density in the finite lattice was then obtained, and coexistence curves with a region of anomaly in the density presented in a phase diagram. In the end, a phenomenological theory for the system density is proposed and compared to the previous results.
\end{abstract}


\end{frontmatter}

\section {Introduction}

Although ubiquitous in nature and subject of many studies, water is a substance that is not yet completely understood by scientists \cite{URLChaplin,Ro96,Be99}.
While most liquids contract upon cooling, the specific volume of water starts to increase at ambient pressure when cooled below $T=4 ^oC$ \cite{Wa64,An76}. Besides, within a certain
range of pressures, water also exhibits an anomalous increase of compressibility 
and specific heat upon cooling \cite{Pr87,Pr88,Ha84}. One of the hypothesis for the presence of the anomalies is
that they are  related to a second critical
point between two liquid phases, a low density liquid
(LDL) and a high density liquid (HDL) \cite{Po92}. 

Water is not an isolated case. There are also other examples of  tetrahedrally
bonded molecular liquids such as phosphorus \cite{Ka00}\cite{Mo03}
and amorphous  silica \cite{La00} that also are good candidates for having
two liquid phases. Moreover, other materials such as liquid metals
\cite{Cu81} and graphite \cite{To97} also exhibit thermodynamic anomalies.
Unfortunately,  a coherent and general interpretation of  the low density liquid and high density liquid phases is still missing.  What type of potential would be appropriated for describing the tetrahedrally bonded molecular liquids? In order to address the question of the potential shape appropriated to describe anomalous systems, a few years ago it has been shown that the presence of two liquid phases can be associated with a potential having an attractive part and two characteristic short-range repulsive distances. The smallest of these two distances is associated with the hard core of the molecule, while the largest one with the soft core. In the case of lattice models, the main strategy has been to associate the van der Waals attraction with   the occupational lattice gas variable and    the hydrogen bond disorder with bond \cite{Fr03,Sa96} or site \cite{Sa93b,Ro96b} Potts states. In the former case, coexistence between two liquid phases may follow from the presence of an order-disorder transition, and a density anomaly is introduced \emph{ad hoc} by the addition to the free energy of a volume term proportional to a Potts order parameter. In the second case, it may arise from the competition between occupational and Potts variables introduced through a dependency of bond strength on local density states. In both cases, however, the connection between the number of Potts states and the presence of the density anomaly is not yet understood.

To circumvent this difficulty, recently another type of the associated lattice gas (ALG)  model was proposed \cite{He05a,He05b}.  The main idea is to represent hydrogen bonds through ice variables\cite{Hu83}\cite{attard}\cite{Na91}\cite{Gu00}, so successful in the description of ice \cite{Li67} entropy, for dense systems. In this case, an order-disorder transition is absent.  Competition between the filling up of the lattice and the formation of an open four-bonded orientational structure  is naturally introduced in terms of the ice bonding variables and no \emph{ad  hoc} introduction of density or bond strength variations is needed. This approach bares some resemblance to that of some continuous models\cite{Si98}\cite{Tr99}\cite{Tr02}, which, however, lack entropy related to hydrogen distribution on bonds. Also, the reduction of phase-space imposed by the lattice allows construction  of the full phase diagram, not always possible for continuous  models. Henriques et al \cite{He05a} have found that this model exhibits two liquid phases and a density anomalous region in the pressure-temperature phase diagram. 

The remainder of the paper goes as follows. In sec.(\ref{The Model}), a simplified model is presented and the simplifications regarding its polarity and degrees of freedom explained; the exact finite lattice analysis of the model is shown on sec.(\ref{The Finite Lattice Approximation}); simpler expressions for the density and pressure of the model are proposed in sec.(\ref{Proposed Density}), and our findings summarized in sec.(\ref{Conclusions}).

\section{The Model}\label{The Model}

Many existing models of water-like fluids use polar representations for the hydrogen bonds between molecules \cite{He05b}. As a starting point for our approach, we take the hamiltonian of equation \ref{hamilt1} for a system that allows 18 possible orientational states per particle and presents a phase diagram with liquid and gas phases as well as density anomaly.

\begin{equation} \label{hamilt1} 
\mathcal{H}=(2u-v)\sum\limits_{\langle ij \rangle}^{N} \sigma _{i} \sigma _{j}  +u\sum \limits_{\langle ij \rangle}^{N} \sum \limits_{l=1}^6\sigma _{i} \sigma _{j} \tau _i^l \tau _j^{m_{i,j,l}} (1-\tau _i^l \tau _j^{m_{i,j,l}} )  -\mu \sum \limits_{i}^{N}\sigma_i   .
\end{equation} 

Where $\tau _i^l$ denotes the $l$-arm of the site $i$ that points to the site $j$, and $\tau_j^{m_{i,j,l}}$ is the opposite arm from the site $j$ pointing to arm $l$ of the site $i$, selected by $m_{i,j,l}$, so to guarantee that both arms are allowed to participate in the effective bonding.

Such polar formulation for bond formation requires several evaluations to determine if a bond between two molecules has occurred or not. In that model, the 18 molecular states generate $n^{18}+1$ terms in the evaluation of the partition function in a lattice with $n$ sites, where the ``+1'' term is due to the degenerated empty site states. 

Because of the complexity of all relevant arrangements among particles in models involving long and short-range interactions, mean field approximations approaches (e.g. Bethe-Peierls) may eventually fail to capture some of the system phases, and most results in the literature are obtained from Monte Carlo or Molecular Dynamics simulations \cite{He05b}.

In this work, we propose and discuss a model similar to equation \ref{hamilt1}, modified so to allow analytical calculations by exact state enumeration in an approach similar to that of Wu et al \cite{wu} and Kawabata et al \cite{chikao}. Analyzing the microscopic structure of the outcoming results from previous Monte Carlo simulations, we learned about the geometries of the stable phases and, based on these geometries, we set the dimensions for the smallest lattice that could accommodate them. Then, an exact summation was done, by enumerating all possible states on the finite lattice, obtaining an approximation for the grand-canonical partition function of the system, and validating the proposed model.

According to the term $\sigma _{i} \sigma _{j} \tau _{i} \tau _{j} (1-\tau _{i} \tau _{j} )$ in  equation \ref{hamilt1}, for all the possible combinations of two neighbor arms, resulting bonds are

\begin{equation}
\label{array1} 
\begin{array}{l} {\left\{+-,-+\right\}\to bond} \\ {\left\{++{\rm \; },{\rm \; }--{\rm \; },{\rm \; }00,{\rm \; }0+,{\rm \; }0-\; {\rm \; },{\rm \; }-0{\rm \; },{\rm \; }+0\right\}\to non-bond} \end{array},\\
\end{equation}

so that probability $P(i,j)$ for a bond occurring between two candidate leg states in adjacent sites $i$ and $j$ is

\begin{equation}
\label{array2} 
\begin{array}{c} {P(bond)=2/9}\\ {\mbox{and}} \\ {P(non-bond)=1-P(bond)=7/9.} \end{array}
\end{equation} 

It is important to notice that these are not physical Boltzmann probabilities, so we are ignoring the corresponding energy costs for now.

To reduce the number of microstates and facilitate the evaluations of the grand partition function, we also did two additional simplifications. The first one has to do with considering the effectiveness of the bonding directly in the hamiltonian, instead of each molecular ``leg'' polarity.

\begin{figure}[htb!]
\begin{center}
\includegraphics[ bb=0bp 0bp 182bp 92bp, clip, scale=1.0]{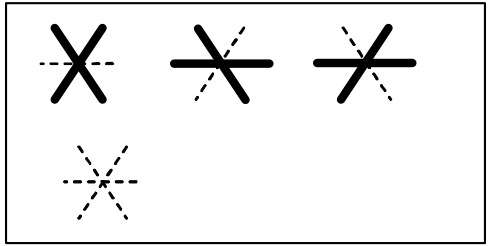}
\end{center}
\caption{Reduced set of states for 2d liquid on the triangular lattice. Lattice sites containing molecule extrema are represented by a number 1, otherwise by a 0.} \label{image1}
\end{figure}

If we change the states for each $\tau$ ``leg'' from $\{-1,0,1\}$ to just $\{0,1\}$, the system hamiltonian can be rewritten, changing the long-range term from $\sigma _{i} \sigma _{j} \tau _{i} \tau _{j} (1-\tau _{i} \tau _{j} )$ to just $-2\sigma _{i} \sigma _{j} \tau _{i} \tau _{j} $. The original model geometry, consisting of four legs allowed to form bonds and two empty links in the triangular lattice was kept unmodified, producing the new set of only 3 states, which are basically $\frac{2}{3}\pi$ ($120^{\circ}$) rotations of the original $X$-shaped structure that represent molecules on the triangular lattice, as can be seen in figure \ref{image1}.


The new probabilities  for a bonding to occur, $P'$, still ignoring the corresponding energetic costs, are

\begin{equation}
\label{array3} 
\begin{array}{c}{
P'(bond)=P(1,1)=(4/12)(4/12)=(16/144)  = 1/9}
\\{\mbox{and}}\\
{P'(non-bond)=1-P'(bond)=P(1,0)+P(0,1)+P(0,0)=8/9.}\\
\end{array}
\end{equation}

As it can be seen from equation \ref{array3}, the new probabilities of randomly generating two connected adjacent bonds, $P'=1/9$, compared to the original $P=2/9$ (equation \ref{array2}), still is much smaller than the probability of not having any bonding, allowing the expectation of a behavior similar to that observed in the original polar model.

To conclude our set of proposed modifications, we will show that it is also possible to get rid of the occupation variable $\sigma$ by defining the relation 

\begin{equation} \label{GrindEQ__2_} 
\sigma =\frac{1}{4} \sum _{i=1}^{6}\tau_{i}^2,   
\end{equation} 

that, applied in the case of apolar representations, allows the replacement of $\tau_{i}^2$ by $\tau_{i}$. This new constraint can be interpreted as \textit{``if there is a leg somewhere in the lattice $\tau$, there must also be a particle in the corresponding site $\sigma$.''} or, considering water molecules without isolated hydrogen or oxygen atoms, a more physical interpretation is \textit{``if an hydrogen atom is found, then there is at least one oxygen atom attached to it''}. The hamiltonian of equation \ref{hamilt1} then becomes

\begin{equation} \label{hamilt2} 
\begin{array}{ccc} {\mathcal{H}'=(2u-v)\sum\limits _{\langle ij \rangle}^{N} \left[  \frac{1}{16} \sum\limits _{l=1}^{6}(\tau _i^l)^2  \sum\limits _{l=1}^{6}(\tau _j^l)^2 \right ]} \\ {+u \sum\limits _{ij}^{N}\left[\frac{1}{16} \sum\limits _{l=1}^{6}(\tau _i^l)^2  \sum\limits _{l=1}^{6}(\tau _j^l)^2  \sum\limits _{l=1}^{6} \left( \tau _i^l \tau _j^{m_{i,j,l}} \left[1-\tau _i^l \tau _j^{m_{i,j,l}} \right]\right) \right ].} \end{array}
 \end{equation}

With this, the $\tau$ variables now carry all information about both short-range(van der Waals) and long-range(hydrogen bonding) interactions. The new expression has a larger number of terms, but a close inspection will reveal that such terms are rather simple, with the advantage of having just one variable to represent both types of degrees of freedom. 

\section{The Finite Lattice Approximation}\label{The Finite Lattice Approximation}

In order to obtain an expression for the density of the system, the grand partition function was approximated evaluating

\begin{equation} \label{soma} 
\Xi(z,\beta)=\sum_{\langle mnop \rangle}{z^N e^{-\beta E_{mnop}}},
\end{equation} 

by enumerating and summing it over all possible configurations \cite{wu,chikao}, with $\langle m,n,o,p \rangle$ denoting the state of the system and $E_{mnop}$ its energy over the $2 \times 2$ lattice with periodic boundary conditions.

Subsequently, an expression for the density of the system was obtained, and a phase diagram with coexistence curves and a maximum of density was generated. 

\begin{figure}[htb!]
\begin{center}
\includegraphics[ bb=0bp 0bp 219bp 214bp, clip, scale=0.7]{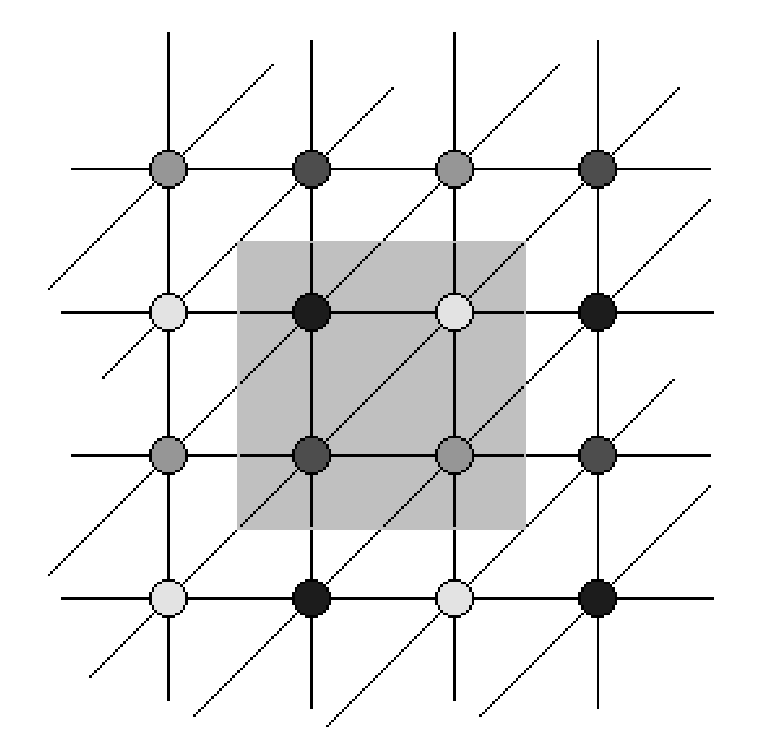}
\end{center}
\caption{Schematics of the four-site lattice used over gray-shaded background. Shaded sites outside the gray square correspond to the site inside the square with the same shading, indicating the periodic connectivity used. } \label{image3}
\end{figure}

As it was presented, the partition function was obtained by expanding the hamiltonian of equation \ref{hamilt1} on the lattice sites and summing over all states. This approach was possible only because the search space in this case was severely reduced due to the simplifications applied, allowing the further use of a software tool to perform the symbolic summations. 
The resulting expression,

\begin{equation} \label{Z3estados} 
\begin{array}{ll} \Xi_{3st} =3 \left( 27 + 108 e^{\beta  \mu }+72 e^{2 \beta  (v+\mu)}+90 e^{ 2 \beta (-2 u+v+\mu )}+4 e^{3 \beta  (2 v+\mu )}+8 e^{3 \beta  (-4 u+2 v+\mu )} \right. \\ + 10 e^{4 \beta  (-4 u+3 v+\mu )}+16 e^{4 \beta  (-3 u+3 v+\mu)} +e^{4 \beta  (-2 u+3 v+\mu )}+60 e^{\beta  (-8 u+6 v+3 \mu )} \\ \left.  +36 e^{\beta  (-4 u+6 v+3 \mu )} \right),
\end{array}
\end{equation} 

 combined with the relation  

\begin{equation} \label{rhoderiv} 
\rho =\frac{1}{\beta V} \frac{\partial \ln \Xi}{\partial \mu } =\frac{1}{\beta V} \frac{1}{\Xi} \frac{\partial \Xi}{\partial \mu } ,
\end{equation}
 
gives the normalized expected particle density,

\begin{equation} \label{rho3estados} 
\begin{array}{ll}
\rho_{3st}=\left[ 9 e^{(2 (6 u+v+\mu ))\beta} \left(5+4e^{(4 u)\beta}\right)+27 e^{(16 u+\mu)\beta}+3 e^{(4 u+6 v+3 \mu )\beta}\left(2+e^{(4 u)\beta}\right) \right. \\ \left(1+7e^{(4 u)\beta}+e^{(8u)\beta}\right) + \left. e^{(4 (3 v+\mu ))\beta} \left(10+16e^{(4 u)\beta}+e^{(8 u)\beta}\right)  \right] \times \\ \left[18e^{(2 (6 u+v+\mu ))\beta} \left(5+4 e^{(4u)\beta}\right)+4 e^{(4 u+6 v+3 \mu )\beta}\left(2+e^{(4 u)\beta}\right) \left(1+7e^{(4 u)\beta}  +e^{(8u)\beta}\right) \right. \\ \left. +e^{(4 (3 v+\mu ))\beta} \left(10+16e^{(4 u)\beta}+e^{(8 u)\beta}\right)+27e^{(16 u)\beta} \left(1+4 e^{(\mu)\beta}\right)\right]^{-1}.
\end{array}
\end{equation}

Where $\mu$ is the chemical potential, $\beta=1/k_BT$ and $V=L^2$ is the number of sites in the lattice.

\begin{figure}[htb!]
\begin{center}
\includegraphics[ bb=91bp 3bp 759bp 417bp, clip, scale=0.5]{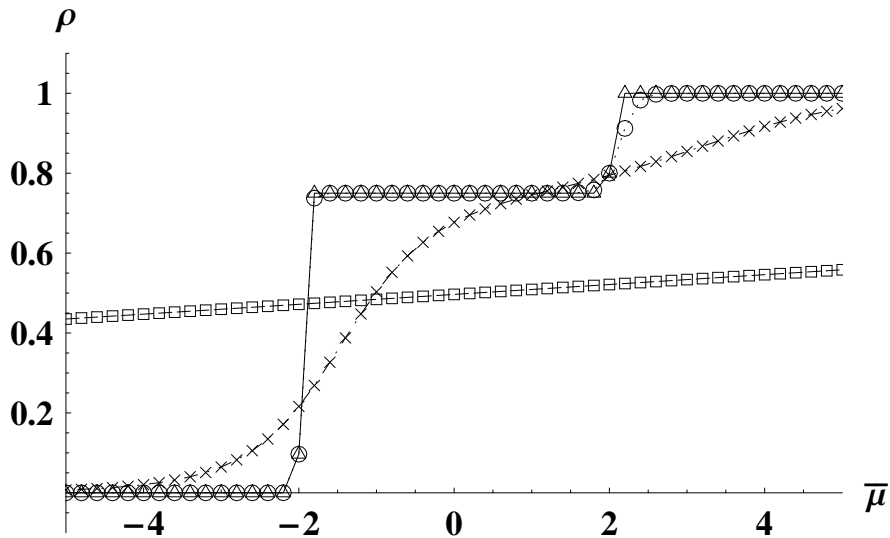}
\end{center}
\caption{Density isotherms $\rho$ against chemical potential $\mu$ from equation \ref{rho3estados}. Triangles correspond to $\bar{T}=10^{-4}$, circles to $\bar{T}=0.1$, X's to $\bar{T}=1.0$, and squares to $\bar{T}=10.0$. It is possible to notice the presence of three stable phases when $\bar{T} < 0.1$. As $\bar{T}$ gets closer to 1, both full and empty phases remain, but the kagomé phase ($\rho=0.75$) gets unstable and vanishes as $\bar{T}$ increases. If we go further towards high $\bar{T}$, we see that $\rho=1$ also doesn't last anymore. }
\label{image5}
\end{figure}

The resulting density for $u=1$, $v=1$, plotted against chemical potential for some values of temperature, depicts curves with the same equilibrium densities already found in more complicated models\cite{He05b}, as can be seen in figure \ref{image5}.

The pressure was then obtained by numerically integrating the Gibbs-Duhem equation:

\begin{equation}\label{gibbsduhem}
SdT-VdP+Nd\mu =0
\end{equation}
and keeping the temperature constant so the first term vanishes. The resulting pressure isotherms can be seen in figure \ref{image5}. From here we start using the overhead bar (e.g. $\bar{T}$) to denote dimensionless variables.

From the pressure curves obtained, it is possible to sketch the phase space of the system. Each level of the pressure steps produces a coexistence curve along $\bar{T}$, in this case, separating regions of gas and low density liquid phases, and also between low and high density liquid phases, respectively.

\begin{figure}[htb!]
\begin{center}
\includegraphics[ bb=91bp 3bp 741bp 405bp, clip, scale=0.5]{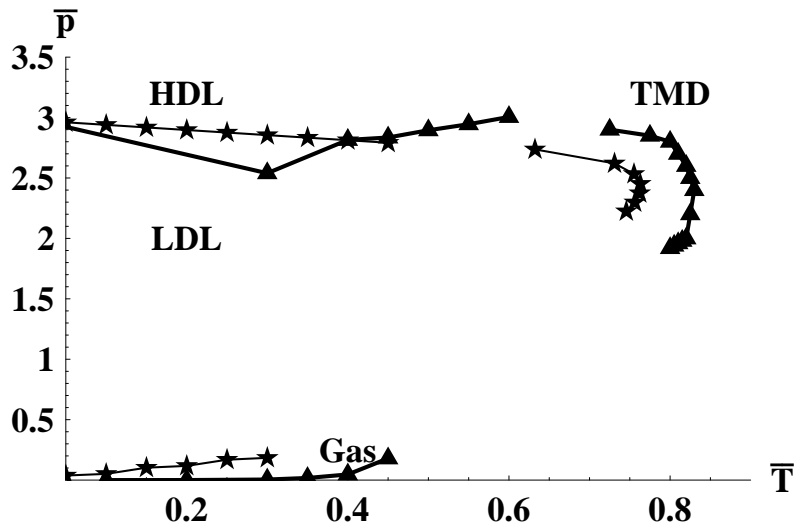}
\end{center}
\caption{Comparison of coexistence curves from Monte Carlo simulations (triangles) \cite{He05a} and finite enumeration (stars) for the transitions Gas-LDL (below), LDL-HDL (above), and maximum of density (right). Parameters $u$ and $v$ are set both to 1.} \label{fase4}
\end{figure}

Investigating the suitable region of the density in $\bar{T}-\mu$ surface, we were also able to identify a region of maxima, as can be seen in the resulting phase diagram of figure \ref{fase4}.

We notice that, even with the known limitations of finite-lattice approximations \cite{Enting1996180}\cite{Landau}, the results from such small lattice dimensions still allowed the occurrence of the same stable phases already shown in \cite{He05a} and \cite{He05b}, resulting in a phase diagram very similar to that found by monte carlo simulations with larger lattice sizes.

\section{Proposed Density}\label{Proposed Density}

Considering the complexity of the expression for the density of the system previously obtained by exact calculations (equation \ref{rho3estados}), the possibility of representing these curves with a simpler expression was considered, so that it could be used to phenomenologically approximate the system behavior and reproduce some of its features - for example, the anomaly of the density. Another important motivation for seeking a simpler solution, is the difficulty to perform an analytical integration in the resulting expression of the exact density to obtain the pressure, for example. With a simplified and integrable representation for the density, it would be possible to obtain expressions for physical observables that provide insights about the real behavior of the system in study. After some simple calculations, we will detail the theoretical hypothesis that motivates this section.

Removing all the inter-particle interaction terms of the hamiltonian, we reduce our model to that of a lattice gas where each particle interacts exclusively with the adsorbant (the lattice), giving 

\begin{equation} \label{hamiltmu}
\mathcal{H}_{ads}=-\mu \sum_{i}^{}  \sigma_i .
\end{equation}

So the grand partition function can be easily obtained by exact state enumeration in a finite lattice,

\begin{equation} \label{Zmu}
\Xi_{ads}= \sum_{\langle \sigma_{mnop} \rangle }^{}  e^{\beta \mu (\sigma_m+\sigma_n+\sigma_o+\sigma_p)} = 81 \left(1+e^{\beta  \mu }\right)^4 \mbox{  (for 3 states)}
\end{equation}

and the density of the system is 

\begin{equation} \label{rhomu}
\rho_{ads}=  \left( 1-\frac{1}{1+e^{\beta  \mu }}\right) = \frac{1}{1+e^{- \beta  \mu }}. 
\end{equation}

The factor $81$ multiplying the partition function in equation \ref{Zmu} arises because the number of states per particle powered to the number of sites is $3^4 =  81$.

The following hypothesis was then raised: Since the exact result obtained by excluding the short and long-range interaction terms from the model has a logistic form, could it be possible to represent the densities of the stable liquid phases of the system as components of a superposition of logistic functions, whose forms are qualitatively compatible with the stable regions of the density $\rho$? In this representation, what characteristics of the original system would be preserved? And what would be the minimum number of terms necessary to provide an ``acceptable'' result from the perspective of structure of the phase diagram?
 
Thus, an initial, phenomenological expression for the density of the system, consisting of a weighted superposition of partial densities based on the equation \ref{rhomu}, was obtained, for the particular case where $ u / v \approx 1$. This specific case greatly simplifies the modelling, and keeps the region where the two liquid phases coexist, because of their importance in the formation of the maximum of density that we intend to study. 

Assuming that the anomalous density of our model can be approximated by a composition of terms similar to the equation \ref{rhomu} for each phase that composes the complete density of the system, the resulting equation is


\begin{equation} \label{expansao} 
\rho (\bar{\mu} ,\bar{T})=\frac{1}{\sum _{i}\rho^s_i  } \sum _{i}^{N_{\rho } }\left[\frac{\rho^s_i }{1+e^{\frac{-\bar{\mu} +\bar{\mu} _{i} }{k_B \bar{T}} } } \right],  
\end{equation} 

where $\rho^s_i$  are the local variations in the stable partial densities for each critical $\mu_c$, and $\mu_i$ is the smallest value of the chemical potential where the $i$-th contribution manifests itself, taking into account the existence of the densities present in smaller chemical potentials.

Analyzing the results from simulations and from the exact analytical solution for the system in the finite lattice with 3 states, we particularize the equation \ref{expansao} for the expression \ref{logistic3} below, 

\begin{equation} \label{logistic3} 
\rho (\bar{\mu} ,\bar{T})=\frac{1}{4} \left[\frac{3}{1+e^{\frac{-\bar{\mu} +\bar{\mu} _{gl} }{k_B \bar{T}} } } +\frac{1}{1+e^{\frac{-\bar{\mu} +\bar{\mu}_{lh} }{k_B T} } } \right],
\end{equation} 

where $\bar{\mu}_{gl} \approx -2 $ and $\bar{\mu}_{lh} \approx 2$ are the chemical potentials of the gas-LDL and LDL-HDL transitions respectively. Such parameteres will depend on the fluid model adopted, and can be estimated by simulation.

The equation \ref{logistic3} is our proposal for the simplified density of a two-dimensional liquid polymorphic fluid, in an expansion consisting of the linear combination of two logistic functions. We then start working with equation \ref{logistic3} for the normalized particle density in order to measure its ability to approximate the behavior of the stable phases of the system. Notice that, in this representation, the parameters $u$ and $v$ are implicit.

\begin{figure}[!htb]
\begin{center}
\includegraphics[ bb=91bp 3bp 759bp 416bp, clip, scale=0.5]{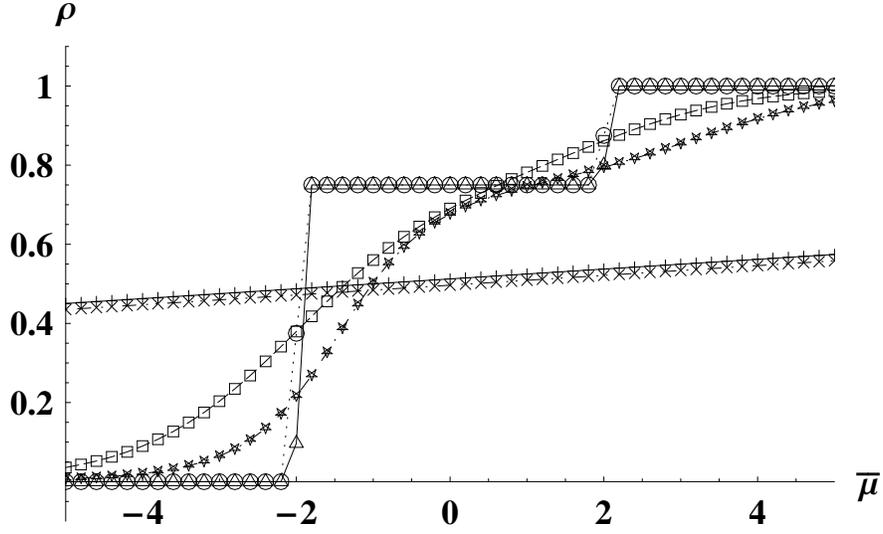}
\end{center}
\caption{Comparison among density curves obtained from the exact solutions in 4 sites for three different temperatures (circles$\to 0.1$ , squares$\to 1.0$ and x's$\to 10.0$), and curves generated by logistic expansions (empty triangles$\to 0.1$, crosses$\to 1.0$ and stars$\to 10.0$). The match between results is higher at low temperatures, decreasing at $\bar{T} \approx 1.0$ and increases again as $\bar{T}$ rises.} 
\label{comparafermi}
\end{figure}

In the figure \ref{comparafermi} we can see the behavior of the density in equation \ref{logistic3}, as a function of the chemical potential $\mu$, for different temperature ranges, in comparison with the exact solution of finite lattice for the model of 3 states in 4 sites previously shown, in identical conditions of $\bar{\mu}$ and $\bar{T}$.

According to figure \ref{comparafermi}, the proposed expresssion for density curves generates results qualitatively similar to that obtained by direct integration of the partition function in the finite lattice. For the case of the specific density that we want to approximate, originated from first-principles physical model, the parameters $\mu_i$ had to be estimated by the values of the chemical potential where the phase transitions happened in monte carlo simulations, taking the low temperature range as a reference.

\begin{figure}[!htb]
\begin{center}
\includegraphics[ bb=145bp 36bp 747bp 408bp, clip, scale=0.6]{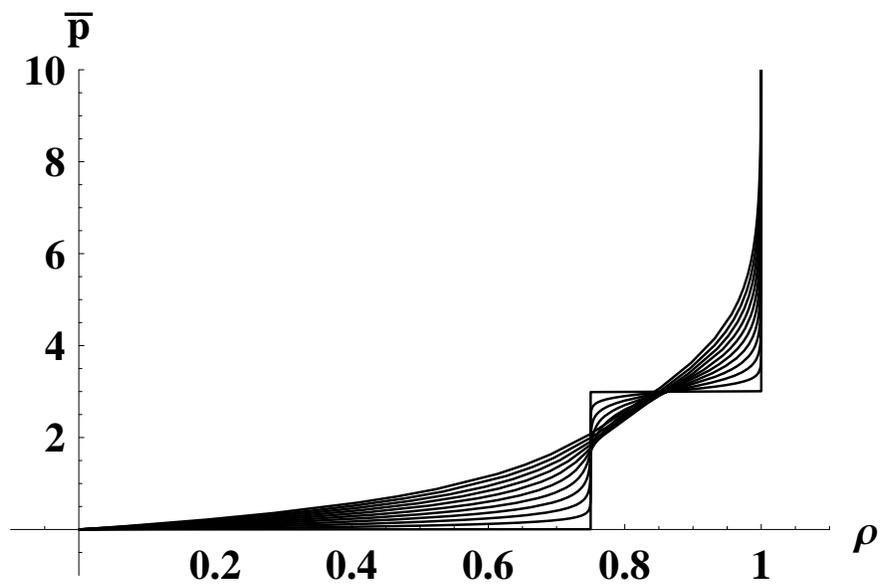}
\end{center}
\caption{Pressure isotherms obtained from equation \ref{prefermi}. The sharper curve corresponds to $\bar{T}=10^{-4}$, until $\bar{T}=10^{0}$ for the smoother one. A close similarity to Monte Carlo results from \cite{He05a} can be noticed.}
\label{pressaofermi1}
\end{figure}

Using once again the Gibbs-Duhem relation, we can obtain an analytical expression for the pressure 

\begin{equation} \label{prefermi} 
\bar{p}=-\frac{\mu_1}{4}+\frac{1}{4} k_B T \left( \log \left[ e^{\frac{\mu}{k_B T}}+e^{\frac{\mu_1}{k_B T}}\right] + 3 \log \left[ 1+e^{\frac{\mu+\mu_2}{k_B T}} \right] \right).
\end{equation} 

 where $-\frac{\mu_1}{4}$ is an integration constant obtained from the condition that the pressure for zero density must also be zero. $\mu_1$ is the chemical potential where the first phase transition , $gas\rightarrow LDL$ occurs, and $\mu_2$ is the chemical potential where the second phase transition , $LDL \rightarrow HDL$ occurs. Isotherms generated from the equation \ref{prefermi} are presented in figure \ref{pressaofermi1}.

\begin{figure}[htb!]
\begin{center}
\includegraphics[ bb=91bp 3bp 805bp 444bp, clip, scale=0.5]{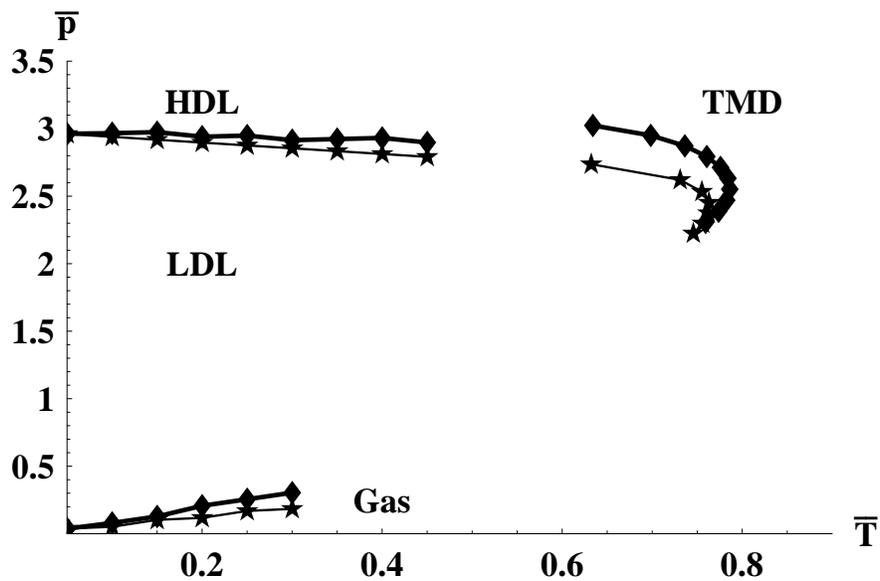}
\end{center}
\caption{Comparison of coexistence curves for the finite enumeration method (stars) and logistic expansion (diamonds). It is possible to see transitions GAS-LDL (below) and LDL-HDL (above). The small curves at right correspond to the maximum of density, captured in both approaches. Notice how most properties of the original model are captured by the approximation proposed in equations \ref{expansao}, \ref{logistic3}, and \ref{prefermi}. } \label{phasecompare}
\end{figure}

The comparison of the resulting phase diagram with a phase diagram obtained by exact integration over a finite lattice can be seen at figure \ref{phasecompare}.

\section{Conclusions}\label{Conclusions}

The  obtained results led us to conclude that finite integrations on a $2 \times 2$ lattice were be able to capture, for a 2-dimensional water-like system, most of the qualitative properties of the system observed in previous Monte Carlo simulations, such as gas, HDL and LDL phases. Although the temperature range of the coexistence lines in our approach is shorter ($\approx20\%$ less), with lower limits, mainly due to the smoothing of transitions because of finite size effects, the phases originally observed in Monte Carlo simulation are still present. The maximum of density also remains in the low-dimensional system, because the quasi-cristalline network in the liquid phase \cite{inhomogeneous}  finds a basis in the lattice with minimal resolution but still enough to occur. It is surprising that such extreme reduction of dimensionality still allows the phase diagram to keep most of its characteristics found in Monte Carlo simulations.

Regarding the proposed (logistic) representation for the system density, most of the features of the phase diagram, obtained previously in results from both Monte Carlo simulations and from finite system approximation, remain present in the phase diagram obtained from the pressure equation integrated from the logistic density. The phase diagram generated using the proposed density also has the anomaly found in previous results from simulations and finite approximation.

Analyzing the $P-T$ phase diagram with the pressure obtained by integrating the logistic density represented in this work by a sum of two more fundamental adsorption curves, we believe that the anomaly phenomenon is the result of two interpenetrating phases, being the less dense a quasi-crystalline phase with vacancies, and the low-density a fluid of non-bonded, loose particles. As temperature rises, vacancies in the quasi-crystalline liquid are filled with free molecules travelling in the fluid, leading to an increase in the density, and as the temperature continues rising, the quasi-crystalline structure breaks, making the density finally starting to decrease.

As a future work, we intend to further investigate these effects both by simulations and analytically in larger systems, in two and three
dimensions, so to get a better understanding about the formation rules underlying the maximum of density in a real, three-dimensional fluid.


\section{Acknowledgements}\label{Acknowledgements}

We want to thank support by the Brazilian science agency CNPq for the provided funding for this study.

\section{References}
\bibliographystyle{unsrt}
\bibliography{mybib2}{}
\end{document}